\documentclass[sigconf]{acmart}

\usepackage{tikz}
\usepackage[pages=some]{background}
\usepackage{hyperref}
\usepackage{graphicx}
\usepackage{svg}

\usepackage{amsmath,amssymb,amsfonts}
\usepackage{listings}
\usepackage{xcolor}
\usepackage{tcolorbox}
\usepackage{algorithmic}
\usepackage{graphicx}
\usepackage{subfigure}
\usepackage{textcomp}
\usepackage{xcolor}
\usepackage{booktabs}
\usepackage{multirow}
\usepackage{makecell}
\usepackage{ulem}
\usepackage{xurl}
\usepackage{titlesec} 

\newcommand{\github}{\url{https://github.com/AutoBench/AutoBench}}
\usepackage{stfloats} 

\newcommand{\acronym}{{AutoBench}}
\newcommand{\acronymeval}{{AutoEval}}
\newcommand{\figname}{{Fig.}}
\newcommand{\refappendix}{Appendix}

\definecolor{vgreen}{RGB}{104,180,104}
\definecolor{vblue}{RGB}{49,49,255}
\definecolor{vorange}{RGB}{255,143,102}

\newcommand{\demosize}{\scriptsize}
\lstdefinestyle{verilog-style}
{
    language=Verilog,
    basicstyle=\demosize{}\ttfamily,
    keywordstyle=\color{vblue},
    identifierstyle=\color{black},
    commentstyle=\color{vgreen},
    numberstyle=\tiny\color{black},
    numbersep=10pt,
    breaklines=true,
    tabsize=8,
    moredelim=*[s][\colorIndex]{[}{]},
    literate=*{:}{:}1
             {[} {{\texttt{[}}}1
             {]} {{\texttt{[}}}1
}

\lstdefinestyle{PYTHON-style}
{
    language=Python,
    basicstyle=\demosize{}\ttfamily,
    keywordstyle=\color{vblue},
    identifierstyle=\color{black},
    commentstyle=\color{vgreen},
    numberstyle=\tiny\color{black},
    numbersep=10pt,
    breaklines=true,
    tabsize=8,
    moredelim=*[s][\colorIndex]{[}{]},
    literate=*{:}{:}1
             {[} {{\texttt{[}}}1
             {]} {{\texttt{[}}}1
}

\lstdefinestyle{python-style}
{
    language=Python,
    basicstyle=\demosize{}\ttfamily,
    keywordstyle=\color{vblue},
    identifierstyle=\color{black},
    commentstyle=\color{vgreen},
    numberstyle=\tiny\color{black},
    numbersep=10pt,
    breaklines=true,
    tabsize=8,
    moredelim=*[s][\colorIndex]{[}{]},
    literate=*{:}{:}1
             {[} {{\texttt{[}}}1
             {]} {{\texttt{[}}}1
}

\newcommand{\codevspacetop}{\vspace{-0.2cm}}
\newcommand{\codevspacebottom}{\vspace{-0.3cm}}

\newcommand{\omitted}{omitted due to space constraints}

\newtcolorbox{databox}[2][]{}
\newtcolorbox{promptbox}[2][]{}

\newcommand{\databoxbegin}[1]{\begin{databox}{#1}\demosize{}\vspace{-0.15cm}}
\newcommand{\databoxend}{\vspace{-0.15cm}\end{databox}}

\newcommand{\indt}{\hspace{0.3cm}}

\copyrightyear{2024}
\acmYear{2024}
\setcopyright{rightsretained}
\acmConference[MLCAD '24]{2024 ACM/IEEE International Symposium on Machine Learning for CAD}{September 9--11, 2024}{Salt Lake City, UT, USA}
\acmBooktitle{2024 ACM/IEEE International Symposium on Machine Learning for CAD (MLCAD '24), September 9--11, 2024, Salt Lake City, UT, USA}
\acmDOI{10.1145/3670474.3685956}
\acmISBN{979-8-4007-0699-8/24/09}

\titlespacing{\section}{0pt}{*0.75}{*0.75}
\titlespacing{\subsection}{0pt}{*0.3}{*0.3}
\titlespacing{\subsubsection}{0pt}{*0}{*1.5}
\begin{document}

\title{\acronym: Automatic Testbench Generation and Evaluation Using LLMs for HDL Design}

\author{Ruidi Qiu}
\affiliation{%
  \institution{Technical University of Munich}
  \city{}
  \country{}
}
\email{r.qiu@tum.de}

\author{Grace Li Zhang}
\affiliation{%
  \institution{TU Darmstadt}
  \city{}
  \country{}
  }
\email{grace.zhang@tu-darmstadt.de}

\author{Rolf Drechsler}
\affiliation{%
  \institution{University of Bremen}
  \city{}
  \country{}
  }
\email{drechsler@uni-bremen.de}

\author{Ulf Schlichtmann}
\affiliation{%
  \institution{Technical University of Munich}
  \city{}
  \country{}
  }
\email{ulf.schlichtmann@tum.de}

\author{Bing Li}
\affiliation{%
  \institution{University of Siegen}
  \city{}
  \country{}
  }
\email{bing.li@uni-siegen.de}


\begin{abstract}
\vspace{-0.1cm}
In digital circuit design, testbenches (TBs) constitute the cornerstone of simulation-based hardware verification.
Traditional methodologies for testbench generation during simulation-based hardware verification still remain partially manual, resulting in inefficiencies in testing various scenarios and requiring expensive time from designers. Large Language Models (LLMs) have demonstrated their potential in automating the circuit design flow. However, directly applying LLMs to generate testbenches suffers from a low pass rate. To address this challenge, we introduce \acronym{}, the first LLM-based testbench generator for digital circuit design, which requires only the description of the design under test (DUT) to automatically generate comprehensive testbenches.
In \acronym{}, 
a hybrid testbench structure and a self-checking system are realized using LLMs.
To validate the generated testbenches, we also introduce 
an automated testbench evaluation framework 
to evaluate the quality of generated testbenches from multiple perspectives.
Experimental results demonstrate that \acronym{} 
achieves a 57\% improvement in the testbench pass@1 ratio compared with the baseline that directly generates testbenches using LLMs. For 75 sequential circuits, \acronym{} successfully has a 3.36 times testbench pass@1 ratio compared with the baseline.
The source codes and experimental results are open-sourced at this link: https://github.com/AutoBench/AutoBench. Artifact DOI: 10.5281/zenodo.13325723.

\end{abstract}




\maketitle
\backgroundsetup{opacity=1, scale=1, angle=0, contents={
\begin{tikzpicture}[remember picture, overlay]
\node[anchor=north east, inner xsep=50pt, inner ysep=10pt] at (current page.north east) {
\href{https://www.acm.org/publications/policies/artifact-review-and-badging-current}{
\includegraphics[width=50pt]{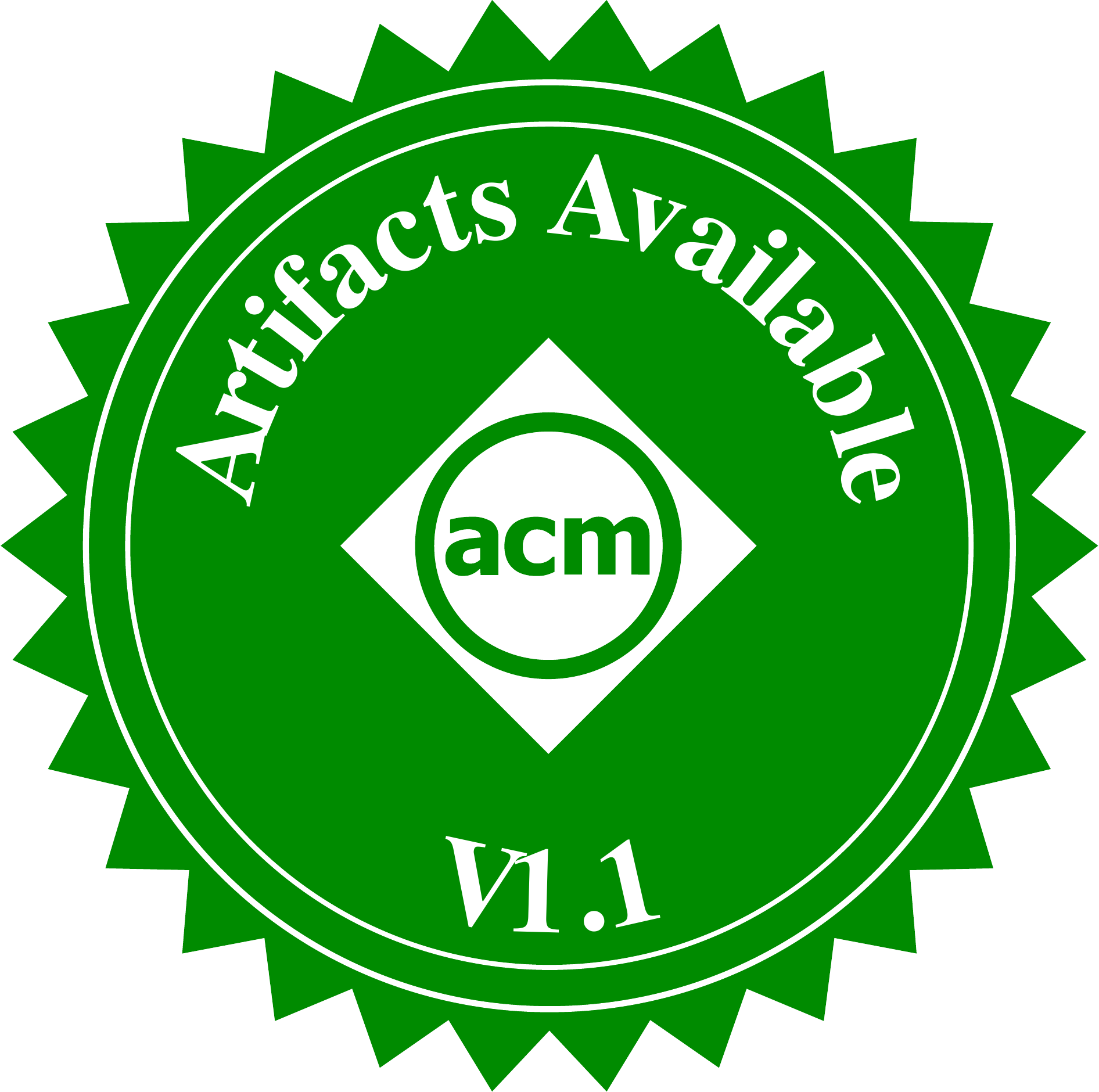}
\includegraphics[width=50pt]{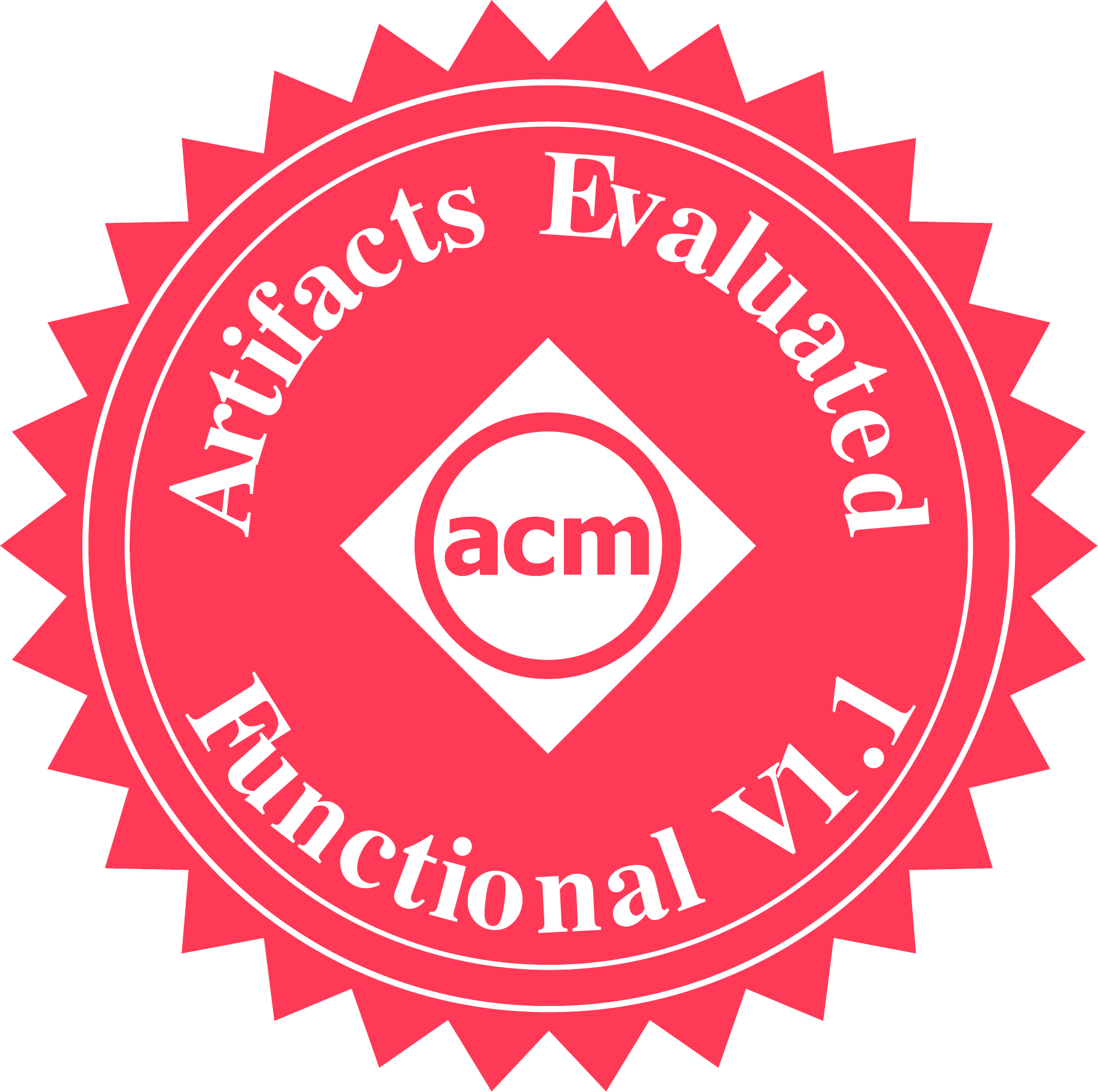}
}};
\end{tikzpicture}
}}
\BgThispage

\newcommand{\secvspace}{\vspace{0cm}}

\section{Introduction}
\vspace{-0.1cm}
\label{sec: introduction}
Simulation-based verification is one of the most common techniques for hardware functional verification 
\cite{scenariocasecoverage_yang}.
This verification is carried out using testbenches to validate the functionality of a DUT.
According to recent studies by the Wilson Research Group \cite{VerificationReport}, around 49\% of a design engineer's time in an IC/ASIC project is spent on conducting verification tasks.
Thus, automating the generation of testbenches is a key point of automating the whole EDA design process of digital circuits.
Previous efforts, such as \cite{murtza2016vertgen, mcellin2022avert}, have sought to automate the code-writing process. But they still need the test stimulus and reference signals designed by engineers. 
Random test case generation approaches, such as constrained random generation (CRG) \cite{kitchen2007stimulus}, have also been proposed to reduce the human effort on test stimulus generation, but the reference signals for checking are still needed from humans.
Thus, these studies could not fully
automated testbench generation, including both the selection of test vectors and the checking of DUT's signals.

Due to the limitations of conventional algorithms, previous research has been unable to achieve complete automation of testbench generation.
With the growing trend of AI applications in hardware design\cite{MLICsurvey}, particularly the advancements facilitated by LLMs, recent studies \cite{chipchat, chipgpt, autochip, verilogeval} on RTL generation using LLMs highlight the proficiency and knowledge of LLMs in digital hardware design.
Additionally, several studies have explored the application of LLMs in the verification process. 
For instance, \cite{llmFVinvariants, autosva2} investigate the potential of LLMs in formal verification, while \cite{zhang2023llm4dv} employs LLMs to generate hardware test stimuli. 
\cite{LLMprocessorverification} offers a case study on LLM-based processor verification. 
Despite these efforts, a systematic approach for LLM-automated testbench generation remains absent, leaving a blank in the automation of the entire simulation-based verification process.

\begin{figure}[t]
    \centering
    \includegraphics[scale=0.9]{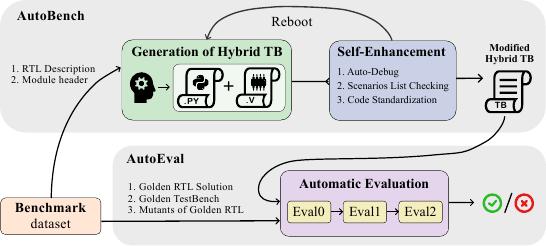}
    \vspace{-0.4cm}
    \caption{Outline of \acronym{} workflow and \acronymeval{} evaluation framework.}
    \label{fig: workflow - outline}
    \vspace{-0.6cm}
\end{figure}
In this paper, an LLM-based testbench generation workflow, \textit{\acronym{}}, is proposed to automatically generate testbenches for simulation-based verification.
A testbench evaluation workflow, \textit{\acronymeval{}}, is introduced to automatically assess the generated testbenches using multiple metrics. 
\acronymeval{} can also be used to test the quality of testbenches generated by other testbench generation frameworks. 
The outline of \acronym{} and \acronymeval{} is illustrated in \figname~\ref{fig: workflow - outline}.
The contributions of this paper are summarized as follows:
\vspace{-0.4cm}

\begin{itemize}
    \item This is the first \textbf{systematic and generic} work to generate Verilog testbenches using LLMs for RTL verification. Unlike previous work on LLM-based verification, our work is validated on a large and widely used dataset. 
    \item A \textbf{hybrid testbench architecture}, including LLM-generated Python code, LLM-generated Verilog code, and Python script, is proposed in our work.
    \item A comprehensive LLM-based \textbf{code generation method} is introduced in our work, including hybrid code generation, scenario checking, code standardization, and LLM-based automatic code debugging and rebooting.
    \item An automatic evaluation framework, \textit{\acronymeval{}}, including a series of general LLM-generated testbench \textbf{evaluating metrics}, is proposed in our work. In addition, a \textbf{dataset} for \acronymeval{} is proposed, which is extended with the help of LLMs from the dataset in previous work \cite{verilogeval}.
    \item The code implementation, the dataset and the experimental results in this work are all \textbf{open-sourced} at \github~.
\end{itemize}
\vspace{-0.1cm}
The rest of the paper is organized as follows. 
In Section \ref{sec: preliminaries}, the background of this work is explained.
In Section \ref{sec: methodology}, the proposed \acronym{} workflow to generate testbenches is explained.
In Section \ref{sec: evaluation}, the proposed \acronymeval{} framework to evaluate the testbench generation framework is elaborated.
Experimental results and conclusions are shown in Section \ref{sec: experimental results} and Section \ref{sec: conclusion}. The detailed demonstrations of \acronym{} forward generation process are shown in the \refappendix.
\secvspace
\section{Background and Motivation}
\label{sec: preliminaries}
\vspace{-0.2cm}
\subsection{Testbench Generation}
\label{sec2 subsec: testbench generation}
\begin{figure}[t]
    \centering
    \includegraphics[scale=0.95]{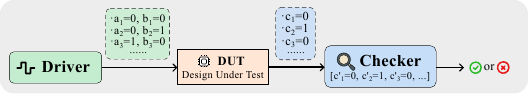}
    \vspace{-0.7cm}
    \caption{The driver and the checker in a TB, assuming the DUT has two input ports ``a'' and ``b'' and one output port ``c''.}
    \label{fig: driver checker}
    \vspace{-0.5cm}
\end{figure}

The conventional testbench design can be split into two parts: the \textit{driver} design and the \textit{checker} design, as shown in \figname~\ref{fig: driver checker}. 
The function of the driver is to generate the input signal vectors (or, in other words, stimuli), drive DUT to generate output signal vectors, and export these vectors to the checker at a proper time.
The checker is responsible for checking if the DUT behaves as required, i.e., checking the output signals from DUT.

An appropriate testbench should be exhaustive and accurate \cite{CoverageMetricsFV}. 
\textit{Exhaustive} implies that the testbench is capable of covering a sufficient number of testing scenarios, while \textit{accurate} denotes that under a given test stimulus, the testbench can provide a correct assessment.

There are several approaches to measure how exhaustive a testbench is, including code coverage, circuit coverage and mutant coverage \cite{CoverageMetricsFV}. The coverage metrics can also be used in test generation \cite{coverage_testgen, coverage_testgen2}.

\subsection{Challenges in LLM-based TB Generation}
\label{sec2 subsec: challenges of LLM tbgen}

LLMs still encounter challenges such as laziness \cite{lazy_llm}, hallucination \cite{hallucinationllm} and insufficient training data on hardware design, making the direct generation of testbenches for complex DUTs ineffective. 
For instance, due to LLM laziness, the generated testbench may end prematurely after listing only a few test cases, resulting in low coverage. 
Additionally, the reference signals required to verify the DUT are not always clear from DUT's specification. 
In complex cases, context-based LLMs may guess incorrect values due to hallucination. 
Furthermore, commercial LLMs such as GPT-4 Turbo are trained more extensively on software codes than hardware codes, indicating a weak performance in hardware contexts. 
To address issues of laziness and hallucination, \acronym{} divides the testbench generation process into multiple stages, applying the idea of chain-of-thought \cite{chainofthought}. 
Additionally, our hybrid testbench structure uses software code, leveraging the advantage of more comprehensive training data in the software domain for a better performance.
\secvspace
\section{Automatic Testbench Generation Framework}
\label{sec: methodology}

\begin{figure*}[t]
    \centering
    \includegraphics[scale=0.9]{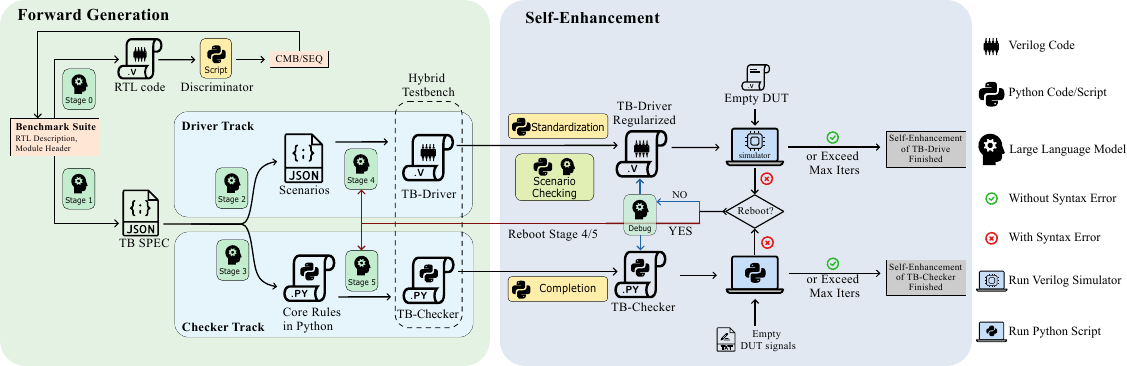}
    \vspace{-0.35cm}
    \caption{\acronym{}: The TB generation workflow in detail.}
    \label{fig: detailed generation workflow}
    \vspace{-0.1cm}
\end{figure*}

In this section, the workflow of \textit{\acronym{}} is explained, including two parts: testbench forward generation in Section \ref{sec3 subsec: generation} and testbench self-enhancement in Section \ref{Sec3 subsec: self-checking}, as is depicted in \figname~\ref{fig: detailed generation workflow}. 
The information we have as input to the \acronym{} workflow includes the DUT's problem description (or, in other words, RTL description) and the module header, as shown in  \figname~\ref{fig: data - benchmark}, where the circuit type (combinational or sequential) is generated by the \acronym{} workflow.
In generating a testbench, the DUT is not available to AutoBench. Otherwise, errors in the DUT may misguide the LLMs so that the resulting testbench may ignore errors.
The following explanation of \acronym{} workflow is accompanied by an example of the testbench generation task \textit{gates100} originally from HDLBits \cite{HDLBits}.
The detailed demonstrations of this testbench generation task include prompts and LLM's responses in each stage, as shown in the \refappendix.

\subsection{Forward Generation Workflow}
\label{sec3 subsec: generation}
The fundamental concept of \acronym{} is to emulate the design process used by human engineers. As outlined in \figname~\ref{fig: driver checker},
the primary generation pipeline is split into two distinct tracks: the driver track and the checker track.
The major steps of the testbench generation workflow in \figname~\ref{fig: detailed generation workflow} are explained in detail below.

\subsubsection{Circuit Type Discriminator}
Digital logical circuits are classified into two types: combinational (CMB) circuits and sequential (SEQ) circuits. 
The circuit type can be derived from the DUT's description, even without DUT's code.
Identifying the circuit type initially allows \acronym{} to provide more precise guidance in subsequent stages to enhance the accuracy of LLM-generated testbenches. 
For instance, testbenches for sequential circuits require stringent time-dependent functions, 
while a checker for a combinational circuit can be a direct function of the current input of the DUT.

To discriminate the types of the DUT, AutoBench generates its own code sample and uses it to decide whether the target DUT is a sequential design or a combinational design.
As shown at the top left corner of \figname~\ref{fig: detailed generation workflow}, given the Benchmark Suite, including the problem description and module header, the LLM is guided to directly generate the RTL code for DUT at \textbf{stage 0}. 
Because the generated RTL is only used for the discriminator in the next step, the correctness of the generated code does not have to be very high as long as there are sufficient features for the later discrimination. 
As the Verilog syntax is strictly restricted by IEEE standard \cite{IEEE_standard_systemVerilog}, the circuit type of an RTL code can be simply determined by special code statements using a Python script. For example, consider a simple instance of \textit{always@(posedge signalx)}, where ``signalx'' represents any signal. This is always a sequential statement.  
The determined circuit type either CMB or SEQ is added to the \textit{Benchmark Suite} to influence the detailed prompts in the following stages, as shown in \figname~\ref{fig: data - benchmark}.

\subsubsection{Verilog Driver Track}
The driver of \acronym{} is a Verilog file, similar to a part of a conventional Verilog testbench file. The difference is that the driver only inputs the test stimuli into the DUT and exports its output. No correctness check on the DUT's output happens in this stage, which is left to the checker part of \acronym{}.
The technical requirements (TRs) for an \acronym{} driver are as follows: 
\textbf{TR1}, the driver's test stimuli must achieve a high coverage rate; 
\textbf{TR2}, the driver should drive the DUT and collect its signals at appropriate times, which is especially important in sequential circuits; 
\textbf{TR3}, the driver should export the DUT's signals in a specified format to be processed by the checker.


After the type of the DUT is identified, the \textit{Benchmark Suite} contains the RTL description and the module header of the DUT, along with the circuit type.  
This information pertains specifically to the RTL code of the DUT rather than the testbench directly, and it lacks certain explicit details, such as a description of the DUT from the perspective of a testbench designer and strategies for designing testbench. 
Without these specific details, the LLM may generate an ineffective testbench, ultimately leading to a verification failure.
To enable the LLM's comprehension of the testbench generation task, we instruct the LLM to summarize a testbench specification in the JSON format from the Benchmark Suite in the next step \textbf{stage 1}, as shown in \figname~\ref{fig: detailed generation workflow}. 
An example of testbench specification generation is shown in \figname~\ref{fig: demo - stage 1}.

Although the design strategies are already contained in the testbench specification, designing the driver simply according to the abstract strategies may cause the laziness of LLMs in choosing test stimuli.
To design test stimuli in the driver with a high coverage to fulfill TR1, a list describing detailed test scenarios is needed to reference the design to the driver code.
Thus, in \textbf{stage 2}, the LLM is directed to generate a list of test scenarios from the information in testbench specification.
A demonstration of stage 2 is shown in \figname~\ref{fig: demo - stage 2}.
Compared to directly generating the testbench, this approach allows the LLM to focus on coverage with split scenarios, thereby achieving a higher coverage ratio.

Subsequently, in \textbf{stage 4} (stage 3 is in Python checker track, which will be introduced later), the LLM ultimately produces the Verilog driver that drives the DUT and collects signals at the correct time points, as is required by TR2, utilizing the testbench specification, test scenarios, and information from the Benchmark Suite. An example of generating a Verilog driver in stage 4 is shown in \figname~\ref{fig: demo - stage 4}.
The instruction for exporting the DUT's signals into a .txt file is also provided in the prompt, thus satisfying TR3.
These signals in the .txt file will be read by the Python checker later to compare the outputs of the DUT and the expected values.

For sequential DUTs, to verify their correctness under a certain test scenario, the testbench needs to access the DUT's signals from not only the current time point but also the previous time points.
Wrong or missing time points to export the signals may mislead the Python checker later and make the checking process fail.
Thus, the generation of Verilog drivers for sequential circuits is more complex than for combinational circuits.
Subsequently, Stage 4 for sequential circuits is divided into two steps. 
The target of the first step is generating the architecture of the driver code and adding the \textit{\$fdisplay} statements to export the DUT's signals at the checking time points. 
The second step is to assert the \textit{\$fdisplay} statements to export signals at the previous time points before checking.
An example of stage 4 for sequential DUTs is demonstrated in \figname~\ref{fig: demo - stage 4 SEQ 1} and \figname~\ref{fig: demo - stage 4 SEQ 2} in the \refappendix.

\subsubsection{Python Checker Track}
With the DUT's signals driven by test stimuli from the Verilog driver, the next step in \acronym{} is to generate the expected output signals with respect to the stimuli 
based on the DUT's description, while the DUT code is still not involved in this track.

Since the checker aims to verify whether the outputs of the DUT with respect to the stimuli are correct, the checker in \acronym{} only needs to generate the expected output signals and compare them with the actual output of the DUT when the testbench is applied. Accordingly, it is not necessary to express the checker in hardware description languages. Instead, we use Python as the language for the checker.
%
%



The utilization of Python code as a testbench checker offers three technical advantages (TAs).
\textbf{TA1 - more appropriate}, because Python, being a higher-level language, provides a greater level of abstraction while ignoring circuit details, making it more suitable for verification tasks than Verilog.
\textbf{TA2 - easier}, since Python benefits from a more extensive dataset than Verilog when LLMs are trained, resulting in a generally superior performance of general conversational language models in Python over Verilog. In addition, Python's extensive standard and third-party libraries make it inherently easier to code than Verilog.
\textbf{TA3 - orthogonal}, because the programming paradigm of Python differs significantly from that of Verilog, which can effectively prevent potential conflicts and errors that may arise from using Verilog codes to verify other Verilog codes. For instance, the designers may make the same error as the DUT's Verilog code when writing the reference RTL's Verilog code for testbench.


As illustrated in the bottom left of  \figname~\ref{fig: detailed generation workflow}, the complexity of the Python checker generation task is mitigated by dividing the Python code generation process into two distinct stages.
The former stage, referred to as \textbf{stage 3}, guides LLM to \textit{translate} the core checking rules of the testbench in natural language into Python, utilizing the testbench specification from stage 1 and other information from Benchmark Suite.
The latter stage, termed \textbf{stage 5}, involves the generation of the complete Python checker code by leveraging the code produced in stage 3, along with the same information used in stage 3, as the prompt to the LLM.
Examples of stage 3 and stage 5 are shown in \figname~\ref{fig: demo - stage 3} and \figname~\ref{fig: demo - stage 5}, respectively.

\begin{figure*}[t]
    \centering
    \includegraphics[scale=0.93]{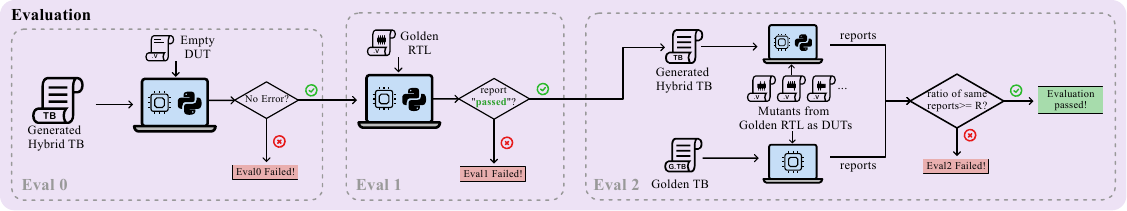}
    \vspace{-0.7cm}
    \caption{\acronymeval{}: The evaluation framework in detail.}
    \label{fig: detailed evaluation workflow}
    \vspace{-0.15cm}
\end{figure*}
\vspace{-0.1cm}
\subsection{Self-Enhancement after generation}
\label{Sec3 subsec: self-checking}
With the generated hybrid testbench, implementing a self-\hspace{0pt}enhancement system is necessary to enhance the testbench.
Note that the self-enhancement system does not involve functional correction for the Python checker because there is no additional data to verify the correctness of the testbench in practice. 

\subsubsection{Code Completion and Standardization}
\label{sec3 subsubsec: completion and standardization}
The Python code from stage 5 is completed by adding a fixed signal interface function, which can read the signals in the .txt file generated by the Verilog driver from stage 4 and send them to the checking function in the Python checker.
For sequential circuits, the Verilog driver must be standardized due to the complexity of the required format, which language models cannot perfectly implement. 
For instance, stage 4's prompt requires the driver to export sequential DUT inputs every clock cycle to ensure that the checker has sufficient information, which is often neglected by LLMs.
The standardization script addresses this by dividing long delays and inserting ``\$fdisplay'' statements. 

\subsubsection{Scenario Checking}\label{sec:scenario_check}
During the integration of test scenarios into the driver code in stage 4, there is a probability that the LLM may omit some scenarios and take shortcuts due to laziness mentioned in Section \ref{sec2 subsec: challenges of LLM tbgen}, thus significantly reducing the testbench's coverage rate. 
To avoid such negligence, scenario checking has been incorporated into our workflow, as is shown in the middle of \figname~\ref{fig: detailed generation workflow}.
Scenario checking uses a Python script to check if all scenarios are included in the Verilog driver code. 
If incomplete, the scenario list from stage 2 and the partial driver code are provided to the LLM to complete the code. 
This process iterates up to a maximum of $I_{SC}$ iterations, set to 3 in Section \ref{sec: experimental results}.

\subsubsection{Auto Debugging and Rebooting}\label{sec:auto_debug_reboot}

Although the functional correctness of the generated testbench cannot be verified, its syntactic correctness can be confirmed using a Verilog simulator and a Python interpreter. 
As illustrated in the upper right of  \figname~\ref{fig: detailed generation workflow}, the driver code is initially simulated with an empty DUT code, which includes only the module header but no content. 
If a syntax error in the testbench arises during this process, the error message and the Verilog code with line markers are provided to the LLM to attempt a resolution using its Verilog knowledge. 
However, the LLM is not capable of rectifying all syntax errors. Consequently, a rebooting system is activated after every $I_{R}$ debugging attempts. 
Upon activation, the system reverts to stage 4 to regenerate the Verilog driver. 
The debugging process for the Python checker is analogous to that of the driver, but the rebooting returns to stage 5 to regenerate the final checker code, as depicted at the bottom middle of  \figname~\ref{fig: detailed generation workflow}. 
Additionally, in some cases, the failure of running the checker code is attributable to the driver code, such as an incorrect signal format transferred to the checker. 
To address this issue, $I_{RV}$ instances of Python errors will trigger the rebooting of the Verilog driver. 
The iterations of debugging and rebooting in total are capped at $I_{D}$ for each type of code. 
In this study, $I_{R}$, $I_{RV}$, and $I_{D}$ are set to 1, 2, and 5, respectively.

\vspace{-0.1cm}
\section{Evaluation Framework for Testbenches}
\label{sec: evaluation}
\vspace{-0.1cm}
\newcommand{\TBunderEval}[1]{TUE#1}
\newcommand{\goldenRTL}[1]{GRTL#1}

A testbench can assess RTL code correctness, but no \textit{hyper-testbench} exists to evaluate the testbench's correctness. 
This study introduces \textit{\acronymeval{}}, a framework with three criteria to assess testbenches under evaluation (TUEs) generated by LLMs. 
\acronymeval{} is versatile, evaluating both hybrid testbenches from the \acronym{} framework and conventional Verilog testbenches. 
The evaluation framework is shown in  \figname~\ref{fig: detailed evaluation workflow}.

The first criterion, \textbf{\textit{Eval0}}, ensures the \TBunderEval{} is syntactically correct and can be compiled. 
The second criterion, \textbf{\textit{Eval1}}, evaluates the preliminary correctness of the \TBunderEval. It reports a status of \textit{passed} if the \TBunderEval{} detects no errors when the golden RTL code is utilized as the DUT.

\begin{table*}[t]
    \caption{Main results of proposed \acronym{} framework.}
    \vspace{-0.4cm}
    \begin{center}

    \begin{tabular}{cc*{3}{|cccc}}
        \toprule
        \multirow{4}*{Group} & \multirow{4}*{Metric} & \multicolumn{4}{c}{Pass@1} & \multicolumn{4}{c}{Pass@5} & \multicolumn{4}{c}{Pass@10}
        \\ \cmidrule(lr){3-6} \cmidrule(lr){7-10} \cmidrule(lr){11-14} & & \multicolumn{2}{c}{Ratio (\%)} & \multicolumn{2}{c}{\#Tasks} & \multicolumn{2}{c}{Ratio (\%)} & \multicolumn{2}{c}{\#Tasks} & \multicolumn{2}{c}{Ratio (\%)} & \multicolumn{2}{c}{\#Tasks}
        \\ 
        \cmidrule(lr){3-4} \cmidrule(lr){5-6} \cmidrule(lr){7-8} \cmidrule(lr){9-10} \cmidrule(lr){11-12} \cmidrule(lr){13-14}  
        & & Ours & Baseline & Ours & Baseline & Ours & Baseline & Ours & Baseline & Ours & Baseline & Ours & Baseline
        \\
        \midrule
        \multirow{3}*{\thead{Total\\(156)}}& Eval2 & 44.81\%   & 28.46\%  & 69.9     & 44.4     & 69.38\%   & 55.78\%  & 108.2    & 87.0     & 76.92\%   & 65.38\%  & 120.0    & 102.0    \\
          & Eval1 & 51.47\%   & 41.73\%  & 80.3     & 65.1     & 81.25\%   & 81.48\%  & 126.8    & 127.1    & 88.46\%   & 92.95\%  & 138.0    & 145.0    \\
          & Eval0 & 95.71\%   & 70.06\%  & 149.3    & 109.3    & 99.97\%   & 98.22\%  & 156.0    & 153.2    & 100.00\%  & 100.00\% & 156.0    & 156.0    \\
        \midrule
        \multirow{3}*{\thead{CMB\\(81)}}  & Eval2 & 62.22\%   & 47.65\%  & 50.4     & 38.6     & 83.39\%   & 80.82\%  & 67.5     & 65.5     & 87.65\%   & 86.42\%  & 71.0     & 70.0     \\
          & Eval1 & 64.81\%   & 58.52\%  & 52.5     & 47.4     & 87.39\%   & 93.44\%  & 70.8     & 75.7     & 93.83\%   & 97.53\%  & 76.0     & 79.0     \\
          & Eval0 & 94.20\%   & 83.58\%  & 76.3     & 67.7     & 99.94\%   & 99.83\%  & 81.0     & 80.9     & 100.00\%  & 100.00\% & 81.0     & 81.0     \\
        \midrule
        \multirow{3}*{\thead{SEQ\\(75)}}  & Eval2 & 26.00\%   & 7.73\%   & 19.5     & 5.8      & 54.25\%   & 28.74\%  & 40.7     & 21.6     & 65.33\%   & 42.67\%  & 49.0     & 32.0     \\
          & Eval1 & 37.07\%   & 23.60\%  & 27.8     & 17.7     & 74.62\%   & 68.58\%  & 56.0     & 51.4     & 82.67\%   & 88.00\%  & 62.0     & 66.0     \\
          & Eval0 & 97.33\%   & 55.47\%  & 73.0     & 41.6     & 100.00\%  & 96.49\%  & 75.0     & 72.4     & 100.00\%  & 100.00\% & 75.0     & 75.0     \\
        \bottomrule
    \end{tabular}
    \label{table: main results}
    \end{center}
\end{table*}

\begin{table}[t]\footnotesize
    \caption{Definitions of Proposed Evaluation Criterion}
    \vspace{-0.4cm}
    \centering
    \begin{tabular}{c|l}
        \toprule
        \textbf{\footnotesize{Type}} & \textbf{\footnotesize{Definition}} \\
        \midrule
        \midrule
        Failed & codes have syntax error \\
        \cmidrule{1-2}
        Eval0 & codes have no syntax error \\
        \cmidrule{1-2}
        Eval1 & codes passed Eval0; report \textit{passed} with the golden RTL code as DUT \\
        \cmidrule{1-2}
        \multirow{2}*{Eval2} & codes passed Eval1; use mutants of golden RTL as DUTs; have the\\ & same report as the golden testbench (\textit{passed} or \textit{failed}) \\
        \cmidrule{1-2}
        Eval2b & similar to Eval2 but use RTL codes generated by LLM as DUTs \\
        \bottomrule
    \end{tabular}
    \label{tab: Evals}
    \vspace{-0.5cm}
\end{table}
However, the aforementioned criteria only identify certain incorrect \TBunderEval{s} and do not address incomplete coverage. 
For example, a \TBunderEval{} that simply reports \textit{no error} for any DUT but without checking the actual signals would always pass both Eval0 and Eval1. 
Thus, an additional coverage-focused criterion is necessary.
Additionally, to ensure objectivity and efficiency, this criterion should be automated. 
Inspired by the mutant coverage metric \cite{CoverageMetricsFV},
we propose \textbf{\textit{Eval2}}, a mutant-based coverage-oriented testbench evaluation criterion.
In Eval2, the DUTs are mutants of the golden RTL code generated by the LLM. 
Both the \TBunderEval{} and the golden testbench (GTB) are run concurrently for each mutant DUT. 
Their results (\textit{passed} or \textit{failed}) are compared. 
The \TBunderEval{} is marked as \textit{success} for that DUT if the results of the \TBunderEval{} and GTB match. 
The proportion of \textit{successes} is the ratio of matches between the 
 \TBunderEval{} and GTB with respect to all the DUT mutants,
which represents the coverage of the TUE.
A testbench passes Eval2 if the Eval2 ratio exceeds a threshold $R$, set to 80\% in Section \ref{sec: experimental results}.

The mutants of the golden RTL codes are generated by the LLM.
The LLM is provided with the golden RTL code and asked to generate $N_{m}$ mutants by making minor modifications evenly on the code. 
To provide the LLM with a better understanding of the code, the RTL description is also provided as a prompt.
In this work, the mutant number $N_{m}$ is set to 10. For some very simple codes, the finally generated mutants could be less than $N_{m}$. 
The definitions of evaluation criteria are summarized in Table \ref{tab: Evals}, where the criterion Eval2b will be discussed in Section \ref{sec4 subsec: performance on llm-generated rtl codes}.

\secvspace

\section{Experimental Results}
\label{sec: experimental results}
\vspace{-0.15cm}
\subsection{Experimental Setup}
\vspace{-0.05cm}
\label{sec4 subsec: experimental setup}

\subsubsection{Software Environment and LLM Model Selection}
In this work, \textit{ICARUS Verilog} \cite{IVerilog} was used as the Verilog simulator, which supports IEEE1800-2012 standards, including system verilog. All the scripts and Python checkers were executed on \textit{Python 3.8.10 64-bit}.
All the experiments were conducted on \textit{gpt-4-turbo-2024-04-09} \cite{OpenAI}. 

\subsubsection{Evaluation Metrics}
Our work is evaluated under multiple evaluation criteria from \acronymeval, including \textit{Eval0}, \textit{Eval1} and \textit{Eval2}, as described in Section~\ref{sec: evaluation} and Table~\ref{tab: Evals}.
The pass@k metric is used, where a testbench generation task is considered to pass an evaluation criterion if $k$ testbenches for the same task are generated and at least one testbench passes this criterion.
To reduce the variance from LLM, the unbiased estimator from \cite{chen2021evaluating} is used:
\begin{equation}
pass@k = \mathop{\mathbb{E}}\limits_{Problems}\left[1-\frac{\tbinom{n-c}{k}}{\tbinom{n}{k}}\right]   
\label{eq: pass@k}
\end{equation}
where $n$ is the total samples we run for each problem and should be as large as possible to guarantee the quality of the testbenches. 
In this work, $n$ was set to 10, considering the cost of using the API of LLMs.


\subsubsection{Dataset}
The dataset used in this study is an extended version of VerilogEval-Human \cite{verilogeval}, an RTL dataset comprising 156 Verilog problems derived from HDLBits \cite{HDLBits}. This extension incorporates RTL mutant codes to facilitate Eval2, as discussed in Section \ref{sec: evaluation}.

\subsection{Main Results}
\label{sec4 subsec: main results}


\subsubsection{Introduction of Baseline}
To demonstrate the performance of the proposed \acronym{}, a comparison experiment between \acronym{} workflow and a baseline workflow was conducted.
The baseline involves a one-step workflow where an LLM receives the \textit{code format}, \textit{problem description}, along with the \textit{DUT header}, and generates the testbench directly. 

\subsubsection{Main Results}
The results of the comparison experiment are shown in Table~\ref{table: main results}.
The first column \textit{group} denotes the circuit type. The Second column \textit{metric} represents the different evaluation criteria introduced in \acronymeval{}. \textit{Ratio} and \textit{\#Tasks} refer to the pass percentage and pass number of test cases, respectively.
In terms of Eval2 pass@1 ratio, which is the most important metric in practice, the third and fourth columns in the fourth row indicate that \acronym{} achieves a 57\% ($\frac{44.81\%}{28.46\%}-1$)  improvement compared with the baseline.
In addition, for sequential tasks, \acronym{} has a 3.36 ($26.00\%\div7.73\%$) times Eval2 pass@1 ratio compared with the baseline (columns 3/4 in row 10). 
Due to the higher complexity and the need to consider timing order, both \acronym{} and the baseline have lower performance for sequential circuits than for combinational circuits. 
With the help of self-enhancement, especially the code standardization for sequential circuits, introduced in Section~\ref{Sec3 subsec: self-checking}, the Eval0 pass@1 ratio of \acronym{} reaches  97.33\% (column 3, row 12), which is a huge improvement compared with 55.47\% of the baseline (column 4, row 12). According to this comparison, 
\acronym{} outperforms the baseline in all aspects. 

\begin{figure}[t]
    \centering
    \includegraphics[scale=0.46]{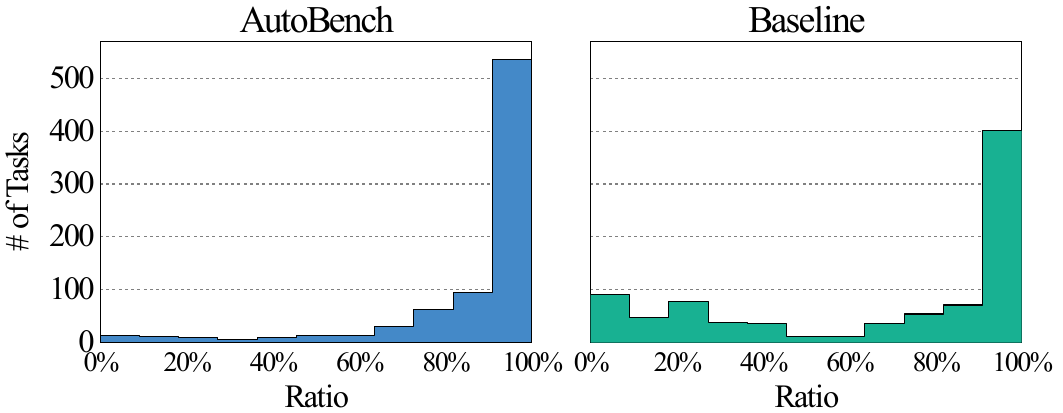}
    \vspace{-0.4cm}
    \caption{The distribution of Eval2 coverages among tasks passing Eval1}
   \vspace{-0.3cm} 
    \label{fig: eval2 distribution}
   \vspace{-0.3cm}
\end{figure}
\subsubsection{Eval2 Coverage Distribution}
In certain instances of Eval1 pass rates (columns 11 and 12 in row 5), \acronym{} performs slightly worse than the baseline ($88.46\% < 92.95\%$). 
This is due to the low testbench coverage of the baseline. 
An extreme example illustrates this: if a testbench reports a ``pass'' for any DUT, its Eval1 pass rate would be 100\%. 
Therefore, Eval2 is required for a comprehensive assessment of testbenches.
Eval2 is a coverage-based criterion and is deemed as \textit{pass} when its ratio is over $80\%$ in this work, as discussed in Section~\ref{sec: evaluation}.
To analyze the detailed coverage distribution of generated testbenches, the ones passing Eval1 from 1560 tasks (156$\times$10 samples) were selected and the distribution of their Eval2 coverage is presented in \figname~\ref{fig: eval2 distribution}. 
Compared with the baseline, Eval2 coverage of \acronym{} is more concentrated on the right side, which means
\acronym{} is more capable of detecting errors in the DUTs than the baseline.

\subsection{Ablation Study of Self-Enhancement}
\label{sec4 subsec: ablation study}

\subsubsection{Impact of Auto-Debug}
\begin{figure}[t]
    \centering
    \hspace{-0.5cm}
    \subfigure[\hspace{0.3cm}Auto-Debug]{
        \includegraphics[scale=0.29]{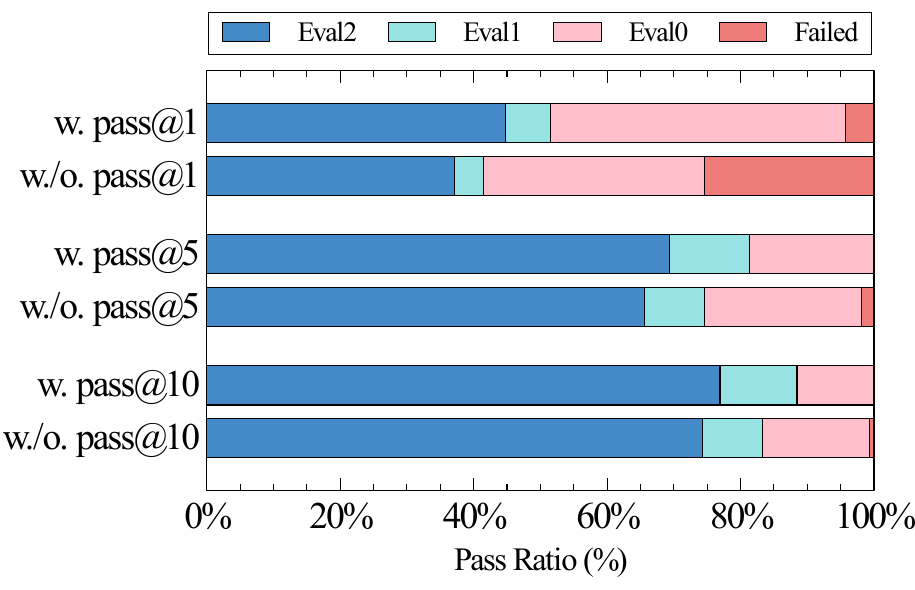} 
        \label{fig: ablation auto-debug}
    }
    \hspace{-0.3cm}
    \subfigure[Scenario Checking]{
        \includegraphics[scale=0.29]{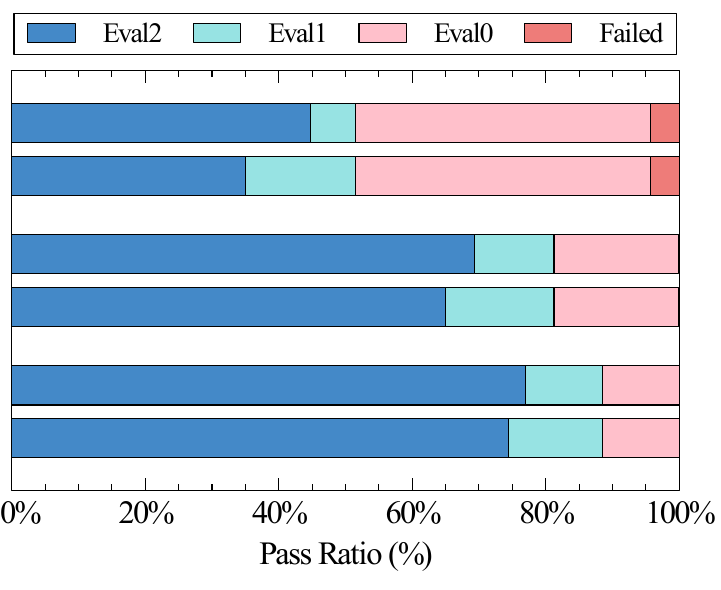}
        \label{fig: ablation scenario checking}
    }    
   \vspace{-0.5cm}
    \caption{Ablation Studies of self-enhancement methods: (a) Auto-Debug and (b) Scenario Checking. The color segments of a bar represent the percentage of tasks that finally passed Eval2/Eval1/Eval0 or failed at Eval0. \textit{w. pass@k} and \textit{w./o. pass@k} represent the pass@k performance with or without the method, respectively.}
   \vspace{-0.6cm}
\end{figure}
To evaluate the impact of auto-debug as described in Section~\ref{sec:auto_debug_reboot},
the performance of \acronym{} without debugging is compared with the original version with debugging of \acronym{} as shown in \figname~\ref{fig: ablation auto-debug}. 
The color segments of a bar in \figname~\ref{fig: ablation auto-debug} represent the percentage of tasks that finally passed Eval2/Eval1/Eval0 or failed at Eval0.
Compared with the version without auto-debug, the full functional \acronym{} has an about 8\% improvement regarding pass@1 Eval2, and 21\% more pass@1 Eval0 ratio.

\subsubsection{Impact of Scenario Checking}

    


To assess the impact of scenario checking described in Section~\ref{sec:scenario_check}, a performance comparison between the presence and absence of scenario checking is shown in \figname~\ref{fig: ablation scenario checking}. 
%
Compared with the version without scenario checking, the fully functional \acronym{} exhibits an approximate 10\% improvement in pass@1 Eval2, along with 4.5\% and 2.5\% enhancements in pass@5 Eval2 and pass@10 Eval2.

\subsection{Performance on LLM-Generated RTL Codes}
\vspace{-0.1cm}
\label{sec4 subsec: performance on llm-generated rtl codes}
\begin{figure}[t]
    \centering
    \includegraphics[scale=0.48]{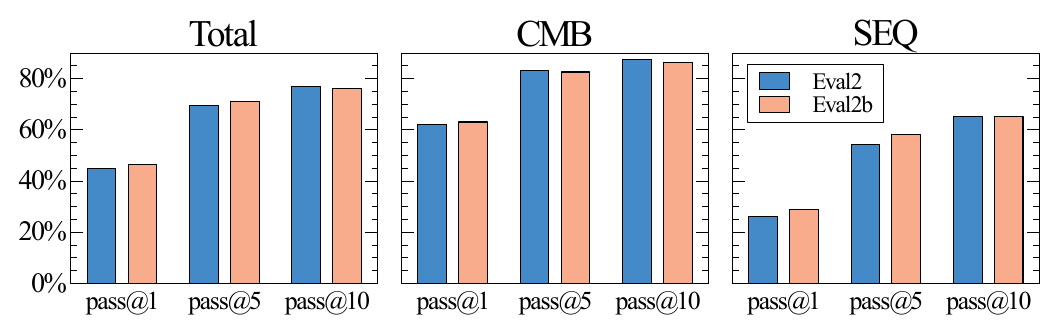}
    \vspace{-0.6cm}
    \caption{Performance on LLM-generated Verilog Codes.}
    \label{fig: eval2b}
    \vspace{-0.6cm}
\end{figure}
The final evaluation metric in prior experiments, Eval2, utilizes the DUTs modified from the golden RTL solution using an LLM. With the increasing prevalence of LLM-assisted RTL code generation, potential challenges may arise when verifying RTL codes generated by the same LLM used in \acronym{}. To test the feasibility of \acronym{} in this context, we replaced the mutant DUTs in Eval2 with LLM-generated RTL codes using the same LLM. This revised metric, termed \textit{Eval2b}, is assessed in \figname~\ref{fig: eval2b}.
The labels at the top of the sub-figures represent the circuit type groups. Two bars in a group on each pass@k metric represent Eval2 (coverage metric on mutant codes from golden RTL) and Eval2b (coverage metric on LLM-generated codes), respectively. These results show that \acronym{} performs similarly or even better on LLM-generated RTL codes compared with Eval2, indicating that our approach is viable for LLM-generated RTL codes.

\subsection{Discussion
}
\vspace{-0.1cm}
Although \acronym{} is designed to generate testbench code without human intervention, it also demonstrates a greater potential for manual correction compared with the baseline. 
The testbench structure from \acronym{} is more organized by test scenarios than the baseline, which usually consists of only simple stimuli and reference signals. 
The core checker codes are written in Python. 
Consequently, even a software engineer can rectify errors that arise in the checker component. 
Furthermore, the scenario-based driver can be easily extended by humans if any scenarios are still missing.
In the future, more self-examination and self-correction approaches will be explored for \acronym.
\secvspace
\section{Conclusion}
\label{sec: conclusion}


In this paper, we propose \acronym{} and \acronymeval{}, the LLM-automated hybrid testbench generation and evaluation frameworks for hardware simulation-based verification. 
\acronym{} demonstrates a 57\%  improvement in the testbench Eval2 pass@1 ratio. For sequential circuits, \acronym{} has 3.36 times Eval2 pass@1 ratio compared with the baseline.
\secvspace
\section*{Acknowledgment}
This work is supported by the Deutsche Forschungsgemeinschaft (DFG, German Research Foundation) -Project-ID 504518248 and by TUM International Graduate School of Science and Engineering (IGSSE).

\secvspace

\let\oldbibliography\thebibliography
\renewcommand{\thebibliography}[1]{%
\oldbibliography{#1}%
\fontsize{8.5pt}{8.5}\selectfont
\setlength{\itemsep}{0.03pt}%
}
\normalem
\bibliographystyle{ACM-Reference-Format}
\bibliography{ref}

\secvspace
\section*{Appendix: Artifact}
\subsection{Abstract}
\label{abstract}
This is the artifact of paper \textit{AutoBench: Automatic Testbench Generation and Evaluation Using LLMs for HDL Design}. All the code implementation, dataset and experimental results have been open-sourced. Please see \github for more detailed information about installation, setup and running.
\subsection{Artifact check-list (meta-information)}

{\small
\begin{itemize}

  \item {\textbf{Model:} OpenAI GPT-4-turbo, GPT-4O, GPT-4O-mini}
  \item {\textbf{Data set:} HDLBits problem set. Already in our project directory.}
  \item {\textbf{Run-time environment:} Linux, Python 3.8 or newer, the latest version of Icarus Verilog (IEEE Std 1800-2012 supported).}
  \item {\textbf{Hardware:} CPU with better performance than i7-11700k.}
  \item {\textbf{Metrics:} Already discussed in paper - syntactic and functional correctness (Eval0, Eval1 and Eval2)}
  \item {\textbf{Output:} code files, conversation files and log files.}
  \item {\textbf{Experiments:} please see the README file in \github.}
  \item {\textbf{How much disk space required (approximately)?:} 4 GB}
  \item {\textbf{How much time is needed to prepare workflow (approximately)?:} 10 minutes.}
  \item {\textbf{How much time is needed to complete experiments (approximately)?:} one problem: 3~10 minutes; 156 problems: 13 hours.}
  \item {\textbf{Publicly available?: }Yes. All the codes, dataset and the experimental results are available to public.}
  \item {\textbf{Code licenses (if publicly available)?:} We sue GPL-3.0 license.}
  \item {\textbf{Workflow automation framework used?:} Yes, this is a totally automatic framework.}
\end{itemize}
}

\subsection{Description}

\subsubsection{How to access}
\label{access}
Please go to \github and follow the detailed guidance. Here is the brief guidance:

\begin{itemize}
    \item Firstly you need to install Python and Icarus Verilog. The version of Python should be at least 3.8. The version of Icarus Verilog should be the latest and perfectly support IEEE Std 1800-2012.
    \item Then please install the required Python libraries. Please see requirements.txt file in github project for the detailed information.
    \item Then please insert your LLM API key into config/key\_API.json before running the project.
    \item Do not forget to change the bin path of iverilog you installed in iverilog\_call.py.
    \item At last, you can run a CMB task demo via "python main.py -d cmb" to make sure you have correctly setup the project. Then you can run "python main.py -d 156" for the whole experiment, or run AutoBench on specific tasks according to the guidance in README.
    
\end{itemize}

\subsubsection{Hardware dependencies}
Check README at \github or section \ref{abstract} Hardware.

\subsubsection{Software dependencies}
Check README at \github or subsection \ref{access}.

\subsubsection{Data sets}

The data set we used is already included in our project directory. You can find it on \url{https://github.com/AutoBench/AutoBench/tree/master/data}.

\subsubsection{Models}

We recommand you use the same LLM as mentioned in paper, i.e. the GPT-4-turbo. We do not guarantee other models will function flawlessly in our work.

\subsection{Installation}

You need to install Python and Icarus Verilog. The version of Python should be at least 3.8. The version of Icarus Verilog should be the latest and perfectly support IEEE Std 1800-2012. Then please install the required Python libraries. Please see requirements.txt file in github project for the detailed information.

\subsection{Experiment workflow}

Check README at \github or subsection \ref{access}.

\subsection{Evaluation and expected results}

The evaluation is automatically performed during running. See Section IV in our paper for the definitions of Eval0/Eval1/Eval2.
The performance should be similar to the original experimental results in \url{https://drive.google.com/drive/folders/1EhG9Ch4vDzMtOsDvoiHthU0OWsZP1xRh?usp=sharing}. 
The Eval 2 pass@1 ratio may fluctuate by up to 5\%.

\subsection{Experiment customization}
We provide multiple running choices. You can run AutoBench with pre-writen configurations or customize your own running. Please see README on \github.
\newpage
\newpage
\section*{Appendix: Code Examples for TB Generation}
\label{appendix}

\newcommand{\demotopspace}{\vspace{-0.4cm}}
\newcommand{\demobottomspace}{\vspace{-0.2cm}}
\begin{figure}[htbp]
    \centering
    \input{codes/stages/benchmark}
    \caption{A demo of Benchmark Suite.}
    \label{fig: data - benchmark}
\end{figure}

\begin{figure}[htbp]
    \centering
    \input{codes/stages/Stage1}
    \caption{A demo of \textit{Stage 1} in \acronym{}. (Further demos are in the following pages.)}
    \label{fig: demo - stage 1}
\end{figure}

\flushbottom
\clearpage

\begin{figure}[htbp]
    \subfigure[stage 2]{
        \vspace{-0.5cm}
        \input{codes/stages/Stage2}
        \vspace{-0.5cm}
        \label{fig: demo - stage 2}
    }
    \subfigure[stage 4]{
        \vspace{-0.5cm}
        \input{codes/stages/Stage4}
        \vspace{-0.5cm}
        \label{fig: demo - stage 4}
    }    
    \vspace{-0.5cm}
    \caption{Demos of the Driver Track: (a) \textit{Stage 2}; (b) \textit{Stage 4}}
\end{figure}


\begin{figure}[htbp]
    \subfigure[stage 3]{
        \input{codes/stages/Stage3}
        \label{fig: demo - stage 3}
    }
    \subfigure[stage 5]{
        \input{codes/stages/Stage5}
        \label{fig: demo - stage 5}
    }    
    \vspace{-0.5cm}
    \caption{Demos of Python Checker Track: (a) \textit{Stage 3}; (b) \textit{Stage 5} }
\end{figure}
\clearpage

\begin{figure}[htbp]
    \centering
    \input{codes/stages/Stage4SEQ1}
    \demotopspace
    \caption{A demo of \textit{Stage 4 part 1} in \acronym{} for sequential DUT, taking sequential task \textit{shift18} as an example.}
    \demobottomspace
    \label{fig: demo - stage 4 SEQ 1}
\end{figure}

\begin{figure}[h]
    \centering
    \input{codes/stages/Stage4SEQ2}
    \demotopspace
    \caption{A demo of \textit{Stage 4 part 2} in \acronym{} for sequential DUT.}
    \demobottomspace
    \label{fig: demo - stage 4 SEQ 2}
\end{figure}

\end{document}